\documentclass[aps,prd,superscriptaddress,showkeys,12pt]{revtex4-2}
\usepackage{inputenc}
\usepackage{amsmath,amsfonts,amssymb}
\usepackage{graphicx, color}
\usepackage{natbib}
\usepackage{setspace}
\usepackage{comment}
\usepackage{orcidlink}
\usepackage{url}  
\usepackage[dvipsnames]{xcolor} 
\usepackage{float}
\usepackage{booktabs} 
\usepackage{hyperref}
\newcommand{\be}{\begin{equation}}
\newcommand{\ee}{\end{equation}}
\newcommand{\bea}{\begin{eqnarray}}
\newcommand{\eea}{\end{eqnarray}}

\begin{document}

\title{A Quaternionic Approach to Teaching 3D Rotations and the Resolution of Gimbal Lock}

\author{Fernando Ricardo González Díaz\orcidlink{0009-0009-9704-1119}}
\affiliation{CICATA-Legaria, Instituto Politécnico Nacional, Ciudad de México, 11500, México.}

\author{Vicent Martinez Badenes\orcidlink{0000-0002-4056-9627}}
\affiliation{Universidad Internacional de Valencia - VIU, 46002, Valencia, Spain.}

\author{Teodoro Rivera Montalvo\orcidlink{0000-0002-6685-0687}}
\affiliation{CICATA-Legaria, Instituto Politécnico Nacional, Ciudad de México, 11500, México.}

\author{Ricardo García-Salcedo\orcidlink{0000-0003-0173-5466}} \thanks{Corresponding author: rigarcias@ipn.mx}
\affiliation{CICATA-Legaria, Instituto Politécnico Nacional, Ciudad de México, 11500, México.}
\affiliation{Universidad Internacional de Valencia - VIU, 46002, Valencia, Spain.}

\date{\today}

\begin{abstract}
Quaternions provide a unified algebraic and geometric framework for representing three-dimensional rotations without the singularities that afflict Euler-angle parametrisations. This article develops a pedagogical and conceptual analysis of the \emph{Gimbal lock} phenomenon and demonstrates, step by step, how quaternion algebra resolves it. Beginning with the limitations of Euler representations, the work introduces the quaternionic rotation operator $v' = q\,v\,q^{*}$, derives the Rodrigues formula, and establishes the continuous, singularity-free mapping between unit quaternions and the rotation group $SO(3)$. The approach combines historical motivation, formal derivation, and illustrative examples designed for advanced undergraduate and graduate students. As an extension, Appendix~A presents the geometric and topological interpretations of quaternions, including their relation to the groups $\mathbb{Q}_8$ and $SU(2)$, and the Dirac belt trick, offering a visual analogy that reinforces the connection between algebra and spatial rotation. Overall, this work highlights the educational value of quaternions as a coherent and elegant framework for understanding rotational dynamics in physics.
\end{abstract}


\maketitle

\section{Introduction}

Quaternions, introduced by Sir William Rowan Hamilton in the nineteenth century, extend complex numbers to four dimensions and provide a robust algebraic and geometric framework for describing rotations and orientations in three-dimensional space. Unlike real or complex numbers, which are confined to one and two dimensions, respectively, quaternions inhabit a four-dimensional space defined by a scalar and a three-component vector. This structure enables a compact and elegant representation of spatial rotations, free from the ambiguities and singularities that afflict other parametrisations such as Euler angles. Their noncommutative nature makes the concept of the quaternion algebra conceptually richer than that of conventional systems. At the same time, their computational efficiency and numerical stability have established them as indispensable tools in computer graphics \citep{vince2011quaternions,taubin20113d}, robotics, aerospace engineering \citep{wie1989quarternion,kristiansen2005satellite,filipe2015adaptive,valverde2018spacecraft}, and physical simulations \citep{rapaport1985molecular,degond2018quaternions,zhu2022review,fu2022accurate}.

A key motivation for using quaternions in rotational dynamics is their ability to avoid the singularities known as \emph{Gimbal lock}, which arise in sequential angle parametrisations when two rotation axes become aligned, resulting in the loss of one rotational degree of freedom. This phenomenon introduces numerical instability in orientation tracking, particularly problematic in aerospace and robotic systems. Unit quaternions provide a compact, globally regular representation of rotation that eliminates these singularities and simplifies the composition, interpolation, and inversion of rotations \citep{kuipers1999quaternions,diebel2006representing}.

Beyond their computational utility, quaternions also offer profound conceptual and educational value. Understanding their historical origin in Hamilton’s quest to generalise complex numbers \citep{lam2003hamilton} fosters an appreciation of how abstract mathematical ideas emerge from the search for more general and unified descriptions of space. Integrating this historical dimension into physics and mathematics education promotes a deeper grasp of the logic and motivation underlying new algebraic structures, making quaternions a fertile topic for interdisciplinary teaching. In this work, we adopt this perspective to design pedagogical materials aimed at undergraduate and graduate students.

Despite their importance in modern physics, many students encounter persistent difficulties when first approaching the mathematics of three-dimensional rotations. Common misconceptions include treating rotation composition as commutative, misunderstanding the relation between Euler angles and physical orientation, and failing to visualize the geometric meaning of quaternionic multiplication. These issues are exacerbated by the coordinate singularities inherent in angle-based systems, which obscure the underlying group structure of spatial rotations. Addressing these obstacles requires instructional resources that connect algebraic formalism with geometric intuition and physical analogy.

Previous contributions to the teaching of rotations—such as the visual explanation of the Dirac belt trick by \citet{silverman1980curious} and the topological demonstrations of $4\pi$ periodicity by \citet{staley2010understanding}—have proven invaluable for introducing students to the nontrivial topology of $SO(3)$. The present work builds upon and extends these efforts by integrating the quaternion formalism into a coherent pedagogical framework that unifies algebraic derivation, computational implementation, and tangible physical models. This integration, rather than new mathematics, constitutes the paper’s primary originality: a reproducible, conceptually transparent approach for teaching three-dimensional rotations through quaternions.

A comprehensive understanding of quaternion algebra involves not only the basic operations of addition, subtraction, multiplication, and division, but also the interpretation of conjugation, norms, inverses, and their relationships with vector operations such as the dot and cross products. These properties make quaternions an excellent didactic vehicle for illustrating the interplay between algebraic structure and geometric transformation \citep{mukundan2002quaternions,barry2016application,bayro2021survey}. In particular, unit quaternions form a continuous group $S^3$ that double-covers the special orthogonal group $SO(3)$, establishing a deep link between quaternion algebra and the topology of spatial rotations.

The article is organised as follows. Section~\ref{sec:quaternion-basics} establishes the algebraic foundations of quaternions, including conjugation, norm, inverse, quaternion–vector identities (dot and cross products), and a homomorphism to complex $2\times 2$ matrices that serves as a bridge to linear–algebraic treatments. Section~\ref{sec:quat-3d-rot} motivates the quaternionic formalism from the limitations of Euler-angle parametrisations, introduces the unit–quaternion axis–angle representation, and frames the \emph{Gimbal Lock} problem in both geometric and computational terms. Section~\ref{subsec:avoid-gimbal-lock} derives the quaternion rotation operator $v'=q\,v\,q^{\ast}$ from first principles, obtains Rodrigues’ formula, and details composition, renormalisation, and SLERP—highlighting the global regularity of unit quaternions ($S^3$) as a double cover of $SO(3)$. Section~\ref{subsec:pedagogical-insights} articulates pedagogical recommendations for teaching 3D rotations with quaternions, and includes a short conceptual bridge to $SU(2)$ and spinors emphasising the $4\pi$ periodicity and the double-cover structure. Section~\ref{sec:teaching-sequence} presents a reproducible classroom sequence (Engage–Explore–Explain–Elaborate–Evaluate) with concrete computational and conceptual tasks. The \textit{Appendix} extends the discussion with a didactic frame-with-strings model linking the Klein group and the quaternion group to physical rotations (including the $4\pi$ untangling), and consolidates the matrix-level correspondence between unit quaternions, Pauli matrices, the isomorphism $S^3\simeq SU(2)$, and its projection onto $SO(3)$.

This integrated approach not only consolidates the mathematical foundations of quaternions but also proposes a pedagogically rich framework for introducing advanced students to abstract algebra and rotational geometry. By combining historical insight, formal derivation, and physical analogy, the work seeks to enhance conceptual understanding and stimulate further exploration of quaternionic structures across physics and engineering.

\section{Exploring Quaternions in Teaching}
\label{sec:teaching-literature}

In recent years, the teaching and learning of quaternions at the university level have gained renewed attention, reflecting their growing relevance in physics, engineering, and computer science curricula. Nevertheless, the academic literature still lacks structured resources that integrate conceptual understanding, computational practice, and pedagogical strategies for effectively introducing quaternions to students.

Several authors have proposed innovative methods for supporting this learning process. \citet{mcdonald2010teaching} suggested a constructive, intuition-based approach linking quaternion operations with rotation matrices, while \citet{rodman2014topics} offered a comprehensive exposition of quaternion linear algebra and its diverse applications. \citet{da2021quaternios} developed educational software to support interactive exploration of quaternion operations, demonstrating the potential of digital tools to facilitate conceptual understanding. In the field of applied mechanics, \citet{markley2008unit} presented a robust method for extracting quaternions from rotation matrices, a procedure now standard in spacecraft attitude determination.

Analogies have also played an essential role in helping students visualise quaternion properties. \citet{staley2010understanding} revisited the Dirac belt trick, explaining its topological significance and its value as a teaching aid to illustrate the $4\pi$ periodicity of spatial rotations. Likewise, \citet{diaz2017fenomeno} expanded on this demonstration, connecting it to the algebraic structure of quaternions and the representation of half-integer spin.

From a broader educational perspective, several authors have stressed the need to contextualise quaternions within physics instruction. \citet{henriksen2014relativity} discussed their relevance to the teaching of relativity and quantum mechanics, where understanding spinor transformations is essential. \citet{bonacci2021teaching} and \citet{montgomery2022introduction} highlighted the importance of motivation and real-world applications in facilitating the comprehension of abstract mathematical structures. A more historical perspective was provided by \citet{kartiwa2023review}, who traced the development of quaternionic differential equations and summarised their mathematical and pedagogical significance.

The present work follows in the spirit of studies such as \citet{familton2015quaternions} and \citet{furui2021understanding}, which emphasise the value of historical and theoretical context in introducing quaternions to physics students. However, our contribution goes further by providing a comprehensive, integrative framework that combines algebraic formulation, geometric interpretation, and pedagogical application. We present detailed numerical examples distinguishing left and right rotations, discuss the physical interpretation of the associated morphisms, and illustrate a three-dimensional rotation using a physical model—a frame with attached strings—that links quaternions, the quaternion group, half-integer spin, and the Pauli matrices \citep{diaz2007topologia, diaz2017fenomeno}.

Moreover, this work addresses one of the most persistent conceptual and computational challenges encountered by undergraduate and engineering students across multiple disciplines: the Gimbal Lock problem \citep{hemingway2018perspectives}. This singularity often arises in courses on mechanics, robotics, aerospace dynamics, and computer graphics, where Euler-angle representations fail to describe specific orientations consistently. Here, we provide a clear and didactic explanation of how unit quaternions eliminate this issue while preserving the physical intuition of rotation. By presenting the problem and its resolution side by side, the discussion becomes not only mathematically rigorous but also pedagogically accessible, allowing students to visualise and understand a difficulty that frequently appears in applied contexts.

This combined theoretical and educational approach aims to bridge the gap between abstract algebraic formulation and practical understanding, providing a reproducible model for integrating quaternion-based rotations into physics and engineering education.

\section{Quaternion algebra: basic definitions}
\label{sec:quaternion-basics}

Quaternions $\mathbb{H}$ extend the complex numbers $\mathbb{C}$, consisting of four basis elements $1, i, j, k$. 
A quaternion $q$ can be expressed as
\begin{equation*}
    q = a + bi + cj + dk,
\end{equation*}
where $a, b, c, d \in \mathbb{R}$, and the imaginary units satisfy the fundamental relations
\begin{equation*}
    i^2 = j^2 = k^2 = ijk = -1.
\end{equation*}
This structure generalises the complex plane to a four-dimensional algebra, where multiplication is associative but not commutative, reflecting the non-commutative nature of spatial rotations.

Let $q_1 = a + bi + cj + dk$ and $q_2 = p + mi + rj + sk$.  
Quaternion addition is performed component-wise:
\begin{equation*}
    q_1 + q_2 = (a + p) + (b + m)i + (c + r)j + (d + s)k.
\end{equation*}
Scalar multiplication of a real number $t \in \mathbb{R}$ by a quaternion $q = a + bi + cj + dk$ is given by:
\begin{equation*}
    t \cdot q = ta + tbi + tcj + tdk.
\end{equation*}

Quaternion multiplication uses the distributive property and the fundamental relations:
\begin{equation*}
    ij = k, \;\; jk = i, \;\; ki = j, \;\; ji = -k, \;\; kj = -i, \;\; ik = -j.
\end{equation*}
Hence, the product is:
\begin{align}
q_1 q_2 &= (a + bi + cj + dk)(p + mi + rj + sk) \nonumber \\
        &= (ap - bm - cr - ds) + (am + bp + cs - dr)i \nonumber \\
        &\quad + (ar - bs + cp + dm)j + (as + br - cm + dp)k. \label{eq:quatprod}
\end{align}

This product can be elegantly written using the dot and cross products of vectors in $\mathbb{R}^3$:
\begin{equation*}
    q_1 q_2 = ap - \mathbf{q}_1 \cdot \mathbf{q}_2 + a\mathbf{q}_2 + p\mathbf{q}_1 + \mathbf{q}_1 \times \mathbf{q}_2,
\end{equation*}
where $\mathbf{q}_1 = (b, c, d)$ and $\mathbf{q}_2 = (m, r, s)$.

The conjugate of a quaternion $q = a + bi + cj + dk$ is
\begin{equation*}
    q^{\ast} = a - bi - cj - dk,
\end{equation*}
and its norm is
\begin{equation*}
    |q| = \sqrt{q q^{\ast}} = \sqrt{a^2 + b^2 + c^2 + d^2}.
\end{equation*}
A quaternion is \emph{unitary} if $|q| = 1$. The inverse of a non-zero quaternion is:
\begin{equation*}
    q^{-1} = \frac{q^{\ast}}{|q|^2}.
\end{equation*}

The matrix representation of quaternions is a powerful tool for understanding their applications in science and engineering. It enables the use of linear algebraic operations to analyse and manipulate spatial rotations.

A structural homomorphism between quaternions and complex $(2\times2)$ matrices can be defined through the mapping $f:\mathbb{H} \rightarrow M_2(\mathbb{C})$:
\begin{equation*}
    f(a+bi+cj+dk) = 
    \begin{bmatrix}
        a+bi & c+di \\
        -c+di & a-bi
    \end{bmatrix}
    =A,
\end{equation*}
where $a, b, c, d \in \mathbb{R}$ and $i, j, k$ are the unit imaginary elements. 
This homomorphism satisfies:
\begin{enumerate}
    \item $f(q_1 + q_2 )=f(q_1 )+f(q_2)$,
    \item $f(r q_1)=r f(q_1)$,
    \item $f(q_1 q_2 )= f(q_1 ) f(q_2)$,
    \item $f(0)=0$,
    \item $f(1)=\mathbf{1}$,
    \item The squared norm of $q$ equals the determinant of $A$: $|q|^2 = a^2+b^2+c^2+d^2 = \det(A)$.
\end{enumerate}

Applying $f$ to the basis elements $\{1,i,j,k\}\in\mathbb{H}$ yields:
\begin{equation*}
    f(1)=
    \begin{bmatrix}
        1 & 0\\
        0 & 1
    \end{bmatrix}=E_1,
    \hspace{1cm}
    f(i)=
    \begin{bmatrix}
        i & 0\\
        0 & -i
    \end{bmatrix}=I_1,
\end{equation*}

\begin{equation*}
    f(j)=
    \begin{bmatrix}
        0 & 1\\
        -1 & 0
    \end{bmatrix}=J_1,
    \hspace{1cm}
    f(k)=
    \begin{bmatrix}
        0 & i\\
        i & 0
    \end{bmatrix}=K_1.
\end{equation*}

Therefore, any quaternion $q=a+bi+cj+dk$ can be expressed as:
\begin{equation*}
    A = aE_1 + bI_1 + cJ_1 + dK_1. 
\end{equation*}

This representation not only provides an elegant algebraic correspondence between $\mathbb{H}$ and $M_2(\mathbb{C})$, but also serves as a practical bridge for students familiar with linear algebra. It allows them to interpret quaternion operations as matrix multiplication, facilitating a smoother transition to understanding three-dimensional rotations and avoiding singularities such as \textit{Gimbal Lock}.

\section{The Gimbal Lock Problem and Quaternions for 3D Rotations}
\label{sec:quat-3d-rot}

Understanding three-dimensional rotations is fundamental in physics, robotics, and computer graphics. While Euler-angle parametrisations are intuitive and historically widespread, they introduce coordinate singularities such as the \emph{Gimbal Lock}. Quaternions provide a compact, numerically stable, and globally regular alternative that we adopt throughout this work \citep{kuipers1999quaternions,diebel2006representing,markley2003attitude,hemingway2018perspectives}.

In two dimensions, rotations are elegantly described by the multiplication of complex numbers. A unit complex number $e^{i\phi}$ rotates a vector by an angle $\phi$ in the plane. Quaternions generalise this concept to three dimensions, extending the algebra of complex numbers into four dimensions. Each unit quaternion encodes a rotation through an axis–angle pair $(\mathbf{u},\theta)$, with $\|\mathbf u\|=1$, as
\begin{equation*}
    q(\theta,\mathbf u)=\cos\!\Big(\tfrac{\theta}{2}\Big)+\sin\!\Big(\tfrac{\theta}{2}\Big)\mathbf u,
\end{equation*}
representing a rotation of angle $\theta$ about the unit axis $\mathbf u\in\mathbb R^3$. This formulation will later allow us to replace Euler angles with a globally regular parametrisation that is free of singularities.

Before introducing the quaternionic formalism, it is instructive to analyse the limitations of Euler-angle parametrisations. In such systems, a general orientation is expressed as three sequential rotations about predefined axes—for instance, $R = R_z(\psi) R_y(\theta) R_x(\phi)$. This approach, while geometrically intuitive, suffers from \emph{Gimbal Lock}, a coordinate singularity that occurs when two of the three rotation axes become aligned, effectively reducing the system’s degrees of freedom from three to two \citep{diebel2006representing, hemingway2018perspectives, markley2003attitude}.

At the singular configuration (\textit{e.g.,} pitch $\theta = \pm 90^\circ$ in the $Z$–$Y$–$X$ convention), the Jacobian of the mapping from Euler angles to orientation loses rank:
\begin{equation*}
    \mathrm{rank}\!\left(\frac{\partial R}{\partial(\phi,\theta,\psi)}\right) < 3,
\end{equation*}
so infinitesimal changes in two angles produce the same orientation, leading to a local loss of invertibility. Physically, this corresponds to the mechanical gimbals becoming coplanar, causing one rotational axis to ``lock'' with another. In aerospace engineering, this loss of control authority can cause catastrophic errors in attitude determination \citep{hemingway2018perspectives}. In computational contexts, it manifests as discontinuities or undefined derivatives during interpolation or integration of orientation data.

Several numerical workarounds exist (e.g., dynamically switching Euler conventions), but these merely relocate the singularity rather than eliminating it. A genuinely global and singularity-free representation requires abandoning angle-based parametrisations in favour of algebraic or geometric structures that remain regular across the entire orientation space.

Quaternions provide precisely such a framework. They extend complex-number algebra into four dimensions, yielding a smooth, single-valued representation of all possible rotations. As we shall demonstrate in the next section, the quaternionic representation $v' = q\,v\,q^*$,
preserves vector norms, composes rotations without ambiguity, and eliminates the coordinate singularities inherent to Euler-angle formulations. Moreover, the topology of unit quaternions—forming the three-sphere $S^3\subset\mathbb R^4$—naturally encodes all orientations in a globally continuous manner \citep{hanson2024visualizingMore}.

Thus, quaternions offer both mathematical elegance and practical robustness. The following section provides a detailed theoretical explanation of how they avoid the \emph{Gimbal Lock} singularity through their algebraic and topological structure.

In the next section, we formalize these ideas by demonstrating, from first principles, how the quaternion algebra intrinsically prevents axis alignment and ensures a smooth, global representation of 3D rotations.

\section{How quaternions avoid the \textit{Gimbal Lock} singularity}
\label{subsec:avoid-gimbal-lock}

As already discussed, a unit quaternion can be expressed as
\begin{equation}
    q(\theta,\mathbf{u}) \;=\; \cos\!\Big(\tfrac{\theta}{2}\Big)\;+\;\sin\!\Big(\tfrac{\theta}{2}\Big)\,\mathbf{u},
    \qquad \|\mathbf{u}\|=1, \nonumber
\end{equation}
which represents a rotation by an angle $\theta$ around the unit axis $\mathbf{u}\in\mathbb{R}^3$.
We identify any $v\in\mathbb{R}^3$ with a \emph{pure quaternion} (zero scalar part). The rotational action
\begin{equation}
    v' \;=\; q\,v\,q^{\*}, \label{qvq}
\end{equation}
preserves the norm and produces the rotation of $v$ around $\mathbf{u}$ by an angle $\theta$.
Using the decomposition $q=q_0+\mathbf{q}$ with $q_0=\cos(\tfrac{\theta}{2})$ and $\mathbf{q}=\sin(\tfrac{\theta}{2})\,\mathbf{u}$,
and the dot–cross product identities in $\mathbb{R}^3$, one obtains the Rodrigues form
\begin{equation}
    v' \;=\; \cos\theta\,v + (1-\cos\theta)\,(\mathbf{u}\!\cdot\! v)\,\mathbf{u} + \sin\theta\,(\mathbf{u}\times v), \label{Eqrodrigues}
\end{equation}
showing that Eq.~(\ref{qvq}) correctly implements the desired rotation \citep{salamin1979, kuipers1999quaternions}.

Unit quaternions form the 3-sphere $S^3\subset\mathbb{R}^4$ and provide a smooth double cover of $SO(3)$ ($q$ and $-q$ encode the same orientation). This global parametrization avoids points where the Jacobian loses rank: no valid orientation lies outside its domain, and no axis alignments appear as in sequential three-angle parametrizations. In practical terms, \emph{every} orientation is represented by a single algebraic object $q$ of unit norm,  without local ambiguity or loss of the third degree of freedom \citep{diebel2006representing, markley2003attitude}.

If $q_1,q_2$ are unit quaternions representing rotations, the composite rotation is given by $q_{\mathrm{comp}} = q_2 q_1$, which preserves the physical order of application (non-commutative) and eliminates the need to manage axis sequences explicitly. Periodic normalization $q\leftarrow q/\|q\|$ controls numerical drift at $\mathcal{O}(1)$ cost, whereas maintaining an orthogonal matrix $R\in SO(3)$ 
typically requires re-orthogonalization \citep{kuipers1999quaternions, markley2008unit}.

Spherical linear interpolation (SLERP) between $q_0$ and $q_1$, defined on $S^3$, 
generates orientation trajectories with constant angular velocity in the proper geometric space,
without crossing singularities or suffering from parametrization distortion:
\begin{equation*}
    \mathrm{slerp}(q_0,q_1;t)\;=\;\frac{\sin((1-t)\Omega)}{\sin\Omega}\,q_0 \;+\; \frac{\sin(t\Omega)}{\sin\Omega}\,q_1,
    \quad \Omega=\arccos(q_0\cdot q_1),
\end{equation*}
with the convention of selecting the short branch ($q_1\leftarrow -q_1$ if $q_0\cdot q_1<0$) \citep{shoemake1985animating}.

In summary, to avoid the \textit{Gimbal Lock} singularity in practical applications, one should: (i) represent orientation exclusively with unit quaternions; (ii) update by multiplication $q\leftarrow \delta q\,q$ (small increments) respecting the order of operations; (iii) periodically renormalize $q$; and (iv) use SLERP for smooth interpolation. This workflow is numerically stable and free of singularities throughout the entire orientation space.

Beyond their computational advantages, unit quaternions also possess a profound geometric and topological interpretation. The set of all unit quaternions, forming the three–sphere $S^3$, is not merely a convenient parameter space but a Lie group that serves as a smooth double cover of the special orthogonal group $SO(3)$. Each spatial orientation corresponds to two antipodal points ${+q,-q}$ on $S^3$, a property that mirrors the double-valued nature of spinor representations in quantum mechanics. This structure establishes the isomorphism $S^3\simeq SU(2)$, where $SU(2)$ denotes the group of $2\times2$ unitary matrices of determinant one. In physical terms, $SU(2)$ provides the natural mathematical language for describing half-integer spin systems, while $SO(3)$ governs the classical rotations of rigid bodies. Thus, the quaternion formalism unifies both under a single geometric picture.

From a pedagogical standpoint, this connection is invaluable: it allows instructors to introduce the concept of spinorial behaviour and $4\pi$ periodicity using a purely geometric argument, before students encounter it formally in quantum mechanics. Visual demonstrations—such as the Dirac belt trick—illustrate how a $2\pi$ rotation leads to a configuration equivalent to $-q$, requiring a full $4\pi$ turn to return to the initial state. This intuitive link between abstract algebraic structure and tangible physical motion provides students with a bridge between classical and quantum representations of rotation. Readers interested in the matrix-level formulation of the mapping $S^3!\rightarrow SU(2)!\rightarrow SO(3)$ are referred to Appendix~A, where these correspondences are presented in full detail.

\section{Pedagogical Insights for Teaching 3D Rotations with Quaternions}
\label{subsec:pedagogical-insights}

In teaching three-dimensional rotations to undergraduate or graduate students in physics and engineering, introducing quaternions offers clear pedagogical advantages over classical angle-based methods. Below, we present guided insights and recommended instructional strategies rooted in both algebraic clarity and experiential learning.

\textbf{1. Build intuition via analogies and transitions.}
Begin by linking familiar two-dimensional rotations in the complex plane to three-dimensional rotations using quaternions. For instance: a unit complex number $e^{i\phi}$ rotates a vector in the plane; analogously, a unit quaternion $q=\cos(\tfrac\theta2)+\sin(\tfrac\theta2)\,\mathbf u$ effects a 3D rotation via $v'=q\,v\,q^*$. This transition helps students see quaternions not as an abstract algebraic curiosity, but as a natural extension of the “complex-number-rotation” notion. \citet{mcdonald2010teaching} presents a constructive method centred on this intuition.

\textbf{2. Use matrix representation as a didactic bridge.}
Introduce the $2\times2$ complex-matrix representation of quaternions (or the equivalent $4\times4$ real form) to show how quaternion algebra preserves linear structure, and how students familiar with linear algebra can visualise quaternion multiplication as matrix multiplication. This strategy links prior knowledge (matrices, eigenvalues) with new content (quaternions) and reinforces the properties of composition, inverses, and unit-norm constraints in a familiar framework.

\textbf{3. Emphasise topological and geometric interpretations.}
Highlight how unit quaternions lie on the 3-sphere $S^3$ and form a smooth double-cover of $SO(3)$. This provides an opportunity to discuss why parametrisation by three sequential angles inevitably leads to singularities (such as gimbal lock), whereas quaternion parametrisation remains globally regular. The anecdotal ''belt trick'' (which ties into the concept of a $4\pi$ rotation returning to identity) can serve both as a visual demonstration and as motivation for students to grasp the ''higher dimensional'' nature of the parameter space \citep{staley2010understanding}.

\textbf{4. Implement hands-on computational activities.}  
Encourage students to code simple rotation routines using quaternions (e.g., in Python, MATLAB, or GeoGebra) with the following tasks:
\begin{itemize}
  \item Given an axis $\mathbf u$ and angle $\theta$, compute $q$ and apply $v'=q\,v\,q^*$ to a set of basis vectors; compare results with the equivalent rotation matrix.
  \item Compose multiple small incremental rotations by quaternion multiplication and observe numerical drift; then apply periodic renormalisation $q\leftarrow q/\|q\|$ and compare.
  \item Perform spherical linear interpolation (SLERP) between two orientations, and visualise the smooth transition on a unit vector; contrast with interpolation using Euler angles, highlighting potential artefacts or singularities.
\end{itemize}
These tasks ground the algebraic formalism in concrete visual and computational practice, reinforcing both understanding and skills.

\textbf{5. Integrate with broader physics/engineering contexts.}  
Link quaternion-based rotation to topics in mechanics, aerospace engineering, robotics and computer graphics. For example:
\begin{itemize}
  \item Attitude representation in spacecraft and inertial systems (unit quaternions vs.\ Euler angles) \citep{Markley2014, forbes2015fundamentals}.
  \item Animation and interpolation of rotations in computer graphics using SLERP \citep{shoemake1985animating}.  
  \item Spinors and their relation to SU(2)-quaternions in quantum mechanics (for more advanced audiences) \citep{staley2010understanding, penrose1984spinors}
\end{itemize}
This contextualisation helps students appreciate the relevance and applicability of quaternions beyond pure mathematics.

Introducing quaternions early in a rotational dynamics or rigid-body kinematics course provides students a robust and unified framework for all subsequent orientation-related topics. Rather than deferring quaternions as an “advanced” aside, embedding them at the heart of rotation instruction encourages deeper conceptual understanding, fewer special-case exceptions (such as gimbal lock), and stronger computational habits.

\subsection{A conceptual bridge: \texorpdfstring{$S^3\simeq SU(2)\twoheadrightarrow SO(3)$}{ } and spinors}

By this correspondence we mean the Lie–group identification between unit quaternions and $SU(2)$, together with the covering homomorphism $\pi: SU(2) \to SO(3)$. Concretely, unit quaternions form the three–sphere $S^3$ and are isomorphic to $SU(2)$; composing this with the projection $\pi$ yields the double cover $S^3\simeq SU(2)\twoheadrightarrow SO(3)$, so that $q$ and $-q$ encode the same spatial rotation via $v\mapsto q\,v\,q^{\ast}$. This section summarises the pedagogical consequences of that structure—spinorial $4\pi$ periodicity, double–valued representations, and the Bloch–sphere picture—while matrix-level details are deferred to Appendix~A.

Unit quaternions form the three–sphere $S^3$ and act on vectors by $v' = q\, v\, q^{\ast}$, yielding all proper rotations in $\mathbb{R}^3$. Conceptually, this structure underlies the well–known double cover $S^3 \!\simeq\! SU(2)\!\twoheadrightarrow\! SO(3)$: each physical orientation in $SO(3)$ corresponds to two antipodal points $\{\pm q\}$ on $S^3$. The “two–to–one’’ mapping explains why quaternions eliminate coordinate singularities while preserving the non-commutativity of finite rotations, and it foreshadows the appearance of spinorial degrees of freedom in quantum theory.

In quantum mechanics, the kinematics of a spin $\tfrac{1}{2}$ system are encoded by state vectors (spinors) in a two-dimensional complex Hilbert space. Physical rotations are represented not in $SO(3)$ but in its double covering $SU(2)$: a spatial rotation of angle $\theta$ around a unit axis $\hat{\boldsymbol n}$ is implemented by the unit function
\begin{equation*}
    U(\hat{\boldsymbol n},\theta)\;=\;\exp\!\Big[-\,\tfrac{i}{2}\,\theta\,(\boldsymbol{\sigma}\!\cdot\!\hat{\boldsymbol n})\Big],
\end{equation*}
where $\boldsymbol{\sigma}$ are the generators (see Appendix A for the matrix formulation). The presence of the factor $\tfrac{\theta}{2}$ is the direct trace of a double cover: a rotation of $2\pi$ in space induces $U=-\mathbb{I}$ on the spinor, and only after $4\pi$ is identity recovered. This property, difficult to visualize with $SO(3)$, becomes natural in $SU(2)$ and has a tangible counterpart in the “Dirac belt trick”, already introduced in this work \citep{staley2010understanding,penrose1984spinors}.

Geometrically, the global phases of a spinor are unobservable, so the pure states of a qubit/spin-$\tfrac{1}{2}$ are represented by points on the Bloch sphere $S^2 \!\simeq\! \mathbb{CP}^1$. The group $SU(2)$ acts transitively on this sphere and projects onto $SO(3)$ onto the axes and angles of rotation in physical space. Thus, the connection “quaternions $\leftrightarrow SU(2)$” provides a direct map between classical rigid-body rotations and transformations of quantum spin states, with a clear correspondence between trajectories in $S^3$ (or in the group $SU(2)$) and orientation curves in $SO(3)$.

Geometrically, the global phases of a spinor are unobservable, so the pure states of a qubit/spin-$\tfrac{1}{2}$ are represented by points on the Bloch sphere $S^2 \!\simeq\! \mathbb{CP}^1$. The group $SU(2)$ acts transitively on this sphere and projects onto $SO(3)$ onto the axes and angles of rotation in physical space. Thus, the connection “quaternions $\leftrightarrow SU(2)$” provides a direct map between classical rigid-body rotations and transformations of quantum spin states, with a clear correspondence between trajectories in $S^3$ (or in the group $SU(2)$) and orientation curves in $SO(3)$.

\section{A short instructional sequence for quaternion-based rotations}
\label{sec:teaching-sequence}

The following teaching sequence is presented as an initial proposal for integrating quaternion concepts into a short physics or engineering module on three-dimensional rotations. A full validation of this proposal would require a more complete instructional design, including assessment instruments and in-class implementation. These steps remain open for future work. Nevertheless, the outline below provides a feasible sequence that can be completed within two 1.5-hour sessions, or a single intensive 2.5-hour workshop.

It is important to note that this sequence is presented as a conceptual and methodological proposal rather than as an empirically tested intervention. Its purpose is to serve as a reproducible framework that instructors can adapt, implement, and subsequently evaluate under controlled educational conditions. The present work thus establishes the theoretical and didactic foundations of the model, leaving its classroom validation and statistical analysis for future research.

The teaching sequence guides students from an intuitive exploration of rotational limitations to a formal understanding and computational application of quaternions, using simple physical demonstrations and computational tools. It follows five natural phases frequently observed in effective physics teaching, although, in principle, no particular pedagogical model is explicitly imposed.

\begin{enumerate}
    \item \textbf{Engage.}
    Begin with a brief demonstration illustrating the limitations of Euler angles and the occurrence of gimbal lock. Start the session by screening a short educational video that clearly shows how the order of rotations affects orientation and how axis alignment leads to a loss of one degree of freedom\footnote{\url{https://www.youtube.com/watch?v=jG2eXkvV0qo} \;(\emph{Rotation Order and Gimbal Lock | 3D Graphics Overview}).}. After viewing, invite students to discuss what they observed and to identify why such a representation may fail in three-dimensional motion. Conclude by posing the question: \emph{Is there a mathematical framework capable of describing any rotation without losing a degree of freedom?} This naturally leads to the introduction of unit quaternions in the following stage.

    \item \textbf{Explore.}  
    Students work in pairs to revisit plane rotations through complex numbers, extending the analogy to three dimensions. They are asked to combine successive small rotations around different axes and to note that the order of application changes the result. As a tactile activity, demonstrate the \emph{Dirac belt trick} or the “box with ribbons” experiment to visualise the $4\pi$ periodicity of rotations and motivate the need for a four-dimensional representation such as quaternions \citep{diaz2007topologia, staley2010understanding, diaz2017fenomeno}.

    \item \textbf{Explain.}
    The instructor gives a 30-minute presentation introducing the quaternion formulation for 3D rotations. The expression $v' = q\,v\,q^{*}$ is derived step-by-step and connected to Rodrigues' formula, emphasizing the geometric significance of quaternion conjugation. Simple diagrams or short animations illustrate how a unit quaternion acts on a vector, conserving its magnitude. Students should note that this unified algebraic representation naturally avoids the gimbal lock singularity mentioned above.

    \item \textbf{Elaborate.}  
    Through a guided computational mini-lab (Python, MATLAB, or GeoGebra), students should do:
    \begin{itemize}
        \item Implement the rotation $v' = q\,v\,q^*$ for a chosen axis–angle pair.
        \item Compare the results with those obtained via rotation matrices.
        \item Perform a smooth interpolation between two orientations using SLERP \citep{shoemake1985animating}.
    \end{itemize}
    Please encourage them to interpret the stability and absence of singularities from a numerical and conceptual perspective.

    \item \textbf{Evaluate.}
    Conclude with a short conceptual and applied evaluation. Suggested tasks include:
    \begin{itemize}
        \item Predicting the result of a composed rotation from given quaternions. \emph{Student task: Given two unit quaternions $q_1$ and $q_2$, compute the composite $q_{\mathrm{comp}}=q_2 q_1$ and apply $v' = q_{\mathrm{comp}}\, v \, q_{\mathrm{comp}}^{*}$ to the basis vectors $\{\mathbf{i},\mathbf{j},\mathbf{k}\}$. Compare your result with the sequential application $v'' = q_2\,(q_1\, v \, q_1^{*})\, q_2^{*}$.}

        \item Explaining why $q$ and $-q$ represent the same orientation. \emph{Student task: Prove that $q$ and $-q$ induce the same rotation by showing $(-q) v  (-q)^{*} = q v  q^{*}$ for any pure quaternion $v$. Provide a one-paragraph explanation of the geometric meaning (double cover $S^3 \to SO(3)$).}

        \item Relating quaternion rotations to practical contexts, such as spacecraft attitude control or 3D animation in computer graphics. \emph{Student task: Select one context (spacecraft attitude or computer graphics). In at most 10 lines, explain why unit quaternions are preferred over Euler angles for (i) composition of rotations, (ii) numerical stability, and (iii) interpolation (SLERP). Include one concrete example or citation.}

        \item Written final reflection (5--7 lines): In your own words, summarise how unit quaternions overcome the limitations of Euler angles and indicate one situation in which using Euler angles might still be acceptable (and why).
    \end{itemize}
\end{enumerate}

This short sequence provides an achievable framework for introducing quaternion-based rotations in a physics or engineering course. It combines conceptual engagement, visual intuition, algebraic derivation, and computational implementation within a compact time frame, fostering both understanding and practical competence. The present work, therefore, introduces this sequence as a theoretically grounded yet exploratory instructional model. While it has not yet undergone empirical classroom testing, its structure is intentionally designed to enable replication and subsequent validation in formal educational environments. Future work could extend this plan by developing assessment rubrics, analysing learning outcomes, and testing its effectiveness through classroom implementation.

\section{Conclusions}
\label{sec:conclusions}

This work offers a unified, pedagogically oriented treatment of quaternions as a framework for representing three-dimensional rotations, highlighting both their mathematical foundations and educational relevance. By tracing the conceptual difficulties associated with Euler-angle parametrisations—particularly the \emph{Gimbal Lock} singularity—and contrasting them with the quaternionic formalism, we have provided a coherent narrative that connects algebraic reasoning, geometric intuition, and computational implementation. The article thus bridges a long-standing gap between abstract formalism and classroom applicability, offering a reproducible model for integrating quaternion-based rotation theory into physics and engineering curricula.

From a pedagogical perspective, the study demonstrates that introducing quaternions through analogies, visual demonstrations, and computational activities can substantially improve conceptual understanding of spatial rotations. The proposed teaching sequence encourages students to move from intuitive observation to formal reasoning and hands-on experimentation. This progression not only facilitates the comprehension of quaternion algebra but also a deeper appreciation of its physical and technological significance across areas such as mechanics, robotics, computer graphics, and quantum physics. By situating quaternions within a broader scientific and historical context, the approach also promotes the development of higher-order skills such as spatial reasoning and abstraction.

Future work should focus on empirically validating the proposed teaching sequence through classroom implementation and assessment of learning outcomes. Designing diagnostic and summative instruments would enable the systematic evaluation of student comprehension and the refinement of the instructional model. Moreover, extending this framework to related domains—such as spinor theory, rigid-body dynamics, and complex-number generalisations—could strengthen the connection between advanced mathematics and physical interpretation. Ultimately, integrating quaternion-based reasoning into undergraduate instruction offers a powerful pathway to cultivate both conceptual coherence and computational fluency in modern physics and engineering education.

\section*{Acknowledgements}

We want to acknowledge the funding provided by the projects SIP20240638, and SIP 20250043 of the IPN Secretaría de Investigación y Posgrado, as well as the FORDECYT-PRONACES-CONACYT CF-MG-2558591 project, which contributed to the development of this work. FRGD sincerely acknowledges Reynaldo González Díaz and José Macario Moreno Calzada for their valuable support.

\bibliographystyle{unsrtnat}
\bibliography{Cuaterniones}

\appendix
\section*{Appendix A. Rotations in Space, Quaternion Groups, and Didactic Models}
\addcontentsline{toc}{section}{Appendix A. Rotations in Space, Quaternion Groups, and Didactic Models}

\subsubsection*{Rotations in space}

Hamilton spent years seeking an algebra for rotations in $\mathbb{R}^3$ based on ordered triples of real numbers \citep{crowe1969history,crowe1994history}. He ultimately realised that achieving a closed and efficient calculus required ordered quadruples—\emph{quaternions}.

Following \citet{lyons2003elementary}, a rotation about the origin in $\mathbb{R}^3$ is specified by an axis (a unit vector) and an angle about that axis. We adopt the convention that rotations are counterclockwise for positive angles and clockwise for negative angles, as viewed from the tip of the axis. This axis–angle specification is not unique: $(\mathbf v,\theta)\sim (k\mathbf v,\theta+2\pi n)$ for any $k>0$ and $n\in\mathbb Z$, and also $(\mathbf v,\theta)\sim(-\mathbf v,-\theta)$.

In linear–algebraic form, rotations are represented by $R\in SO(3)$ (nine parameters subject to six constraints), whereas unit quaternions provide a minimal, numerically stable parametrisation on the three–sphere $S^3$. This economy eliminates coordinate singularities inherent to sequential angle descriptions and yields a single algebraic rule for composition, inversion, and interpolation.

From a pedagogical standpoint, the contrast between the redundancy of matrix parametrisations and the minimality of unit quaternions offers a clear entry point for students to grasp why quaternion algebra is a geometrically efficient language for three–dimensional rotations.

\paragraph{Unit quaternions and the rotation operator.}
Identify any vector $v\in\mathbb{R}^3$ with a \emph{pure quaternion} (zero scalar part). Let a unit quaternion be written as
\begin{equation*}
    q = q_0+\boldsymbol{q} = \cos\!\Big(\tfrac{\theta}{2}\Big)+\sin\!\Big(\tfrac{\theta}{2}\Big) \boldsymbol{u}, \qquad \|\boldsymbol{u}\|=1,
\end{equation*}
and define the rotational action by conjugation $L_q(v)\;=\;q\,v\,q^{\ast}$. Using the decomposition $q=q_0+\boldsymbol{q}$ and the dot–cross identities in $\mathbb{R}^3$ one obtains, for any pure quaternion $v$,
\begin{equation}
    L_q(v)\;=\;\big(q_0^2-\|\boldsymbol{q}\|^2\big)\,v\;+\;2\,(\boldsymbol{q}\!\cdot\! v)\,\boldsymbol{q}\;+\;2\,q_0\,(\boldsymbol{q}\times v), \label{eq:Lq_vector}
\end{equation}
which, after substituting $q_0=\cos(\tfrac{\theta}{2})$ and $\boldsymbol{q}=\sin(\tfrac{\theta}{2})\,\boldsymbol{u}$, yields Rodrigues’ formula
\begin{equation}
    L_q(v)\;=\;\cos\theta\,v\;+\;(1-\cos\theta)\,(\boldsymbol{u}\!\cdot\! v)\,\boldsymbol{u}\;+\;\sin\theta\,(\boldsymbol{u}\times v). \label{eq:Rodrigues_app}
\end{equation}
Two immediate observations follow: (i) $\|L_q(v)\|=\|v\|$, since $|q|=|q^\ast|=1$ for unit $q$ (hence $q^\ast=q^{-1}$); (ii) if $v$ is parallel to $\boldsymbol{u}$, then $L_q(v)=v$ (the rotation axis is invariant). Consequently, $L_q$ is a proper orthogonal map, $L_q\in SO(3)$.

This explicit derivation connects algebraic manipulation with geometric transformation, reinforcing conceptual links between operator formalism and physical rotation.

\paragraph{Linearity.}
For any $a_1,a_2\in\mathbb{R}$ and $v_1,v_2\in\mathbb{R}^3$ (identified as pure quaternions),
\begin{equation*}
    L_q(a_1 v_1 + a_2 v_2)\;=\;a_1\,L_q(v_1)\;+\;a_2\,L_q(v_2).
\end{equation*}
Thus $L_q:\mathbb{R}^3\to\mathbb{R}^3$ is $\mathbb{R}$–linear.

\paragraph{Half–angle (consistency).}
Throughout, we write the unit quaternion as
\begin{equation*}
    q=\cos\!\Big(\tfrac{\theta}{2}\Big)+\sin\!\Big(\tfrac{\theta}{2}\Big)\,\boldsymbol{u}, \qquad \|\boldsymbol{u}\|=1, \label{eq:q_half_angle}
\end{equation*}
so that $L_q$ implements a rotation by \emph{angle $\theta$} about the axis $\boldsymbol{u}$. A formal proof can be found, for example, in \citet{jia2008quaternions}.

\begin{figure}[h]
\centering
\includegraphics[width=7cm,height=6cm]{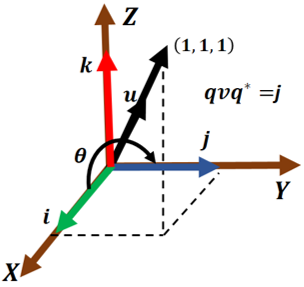}
\caption{The unit quaternion $q=\tfrac{1}{2}(1+i+j+k)$ rotates the basis vector $v=i$ into $j$ for axis $\mathbf{u}=(1,1,1)/\sqrt{3}$ and angle $\theta=2\pi/3$.}
\label{rota3Di}
\end{figure}

\begin{figure}[h]
\centering
\includegraphics[width=7cm,height=6cm]{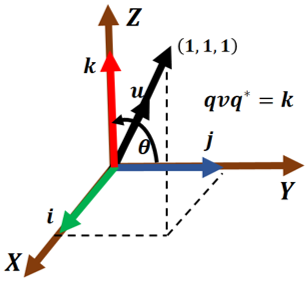}
\caption{For the same unit quaternion $q=\tfrac{1}{2}(1+i+j+k)$, the rotation maps $v=j$ into $k$ under a $2\pi/3$ turn about $\mathbf{u}=(1,1,1)/\sqrt{3}$.}
\label{rota3Dj}
\end{figure}

\begin{figure}[h]
\centering
\includegraphics[width=7cm,height=6cm]{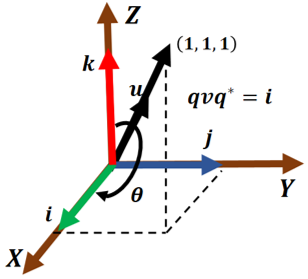}
\caption{The rotation generated by $q=\tfrac{1}{2}(1+i+j+k)$ sends $v=k$ into $i$, completing the cyclic permutation of $(i,j,k)$ about $\mathbf{u}=(1,1,1)/\sqrt{3}$.}
\label{rota3Dk}
\end{figure}

\paragraph{Examples.}
Let $\boldsymbol{u}=\tfrac{1}{\sqrt{3}}(1,1,1)$ and $\theta=\tfrac{2\pi}{3}$. Then
\begin{equation*}
    q=\cos\!\Big(\tfrac{\pi}{3}\Big)+\sin\!\Big(\tfrac{\pi}{3}\Big)\,\boldsymbol{u}
    =\tfrac12\;+\;\tfrac12 i + \tfrac12\,j\;+\;\tfrac12\,k,
\end{equation*}
which is a unit. Using~\eqref{eq:Rodrigues_app} (or direct multiplication via Eq. \eqref{eq:quatprod}) one obtains
\begin{equation*}
    L_q(i)=j,\qquad L_q(j)=k,\qquad L_q(k)=i,
\end{equation*}
\textit{i.e.} a cyclic permutation of the basis directions, as expected (see Figs.~\ref{rota3Di}, \ref{rota3Dj}, \ref{rota3Dk}).

\paragraph{Space–fixed vs.\ body–fixed frames.}
The action $L_q(v)=qvq^\ast$ can be interpreted as rotating the \emph{vector} with respect to a space–fixed frame. Conversely, $L_{q^\ast}(v)=q^\ast v q$ corresponds to rotating the \emph{frame} by $-\theta$ about the same axis, leaving the vector fixed in space.

\paragraph{The group $S^3$.}
The set of unit quaternions $S^3=\{\,q\in\mathbb{H}\;:\;|q|=1\,\}$ forms a (non-abelian) Lie group under multiplication, with identity $1$ and inverses given by $q^{-1}=q^{\ast}$. The map
\begin{equation*}
    \Pi:S^3\longrightarrow SO(3),\qquad \Pi(q)(v)=q\,v\,q^{\ast},
\end{equation*}
is a smooth surjective homomorphism with kernel $\{\pm 1\}$; hence $S^3$ is a double cover of $SO(3)$ and $q$ and $-q$ encode the same physical orientation.

\subsubsection*{Quaternions and rotations of a frame in $\mathbb{R}^3$}

We now focus on a didactic model: rotations of a picture frame and their relationship with discrete groups.

\paragraph{The quaternion group $\mathbb Q$.}
\[
\mathbb Q=\{1,-1,\pm i,\pm j,\pm k\},
\]
with $i^2=j^2=k^2=-1$, $ij=k$, $jk=i$, $ki=j$, and cyclic anticommutativity. The Cayley table (Table~\ref{multgrupo}) summarises the products (adapted from \citealt{weissteinquaternion}).

\begin{table}[h]
\centering
\caption{Cayley table for the quaternion group $\mathbb{Q}_8=\{1,-1,\pm i,\pm j,\pm k\}$.}
\begin{tabular}{|c||c|c|c|c|c|c|c|c|}
\hline
$\cdot$ & $1$ & $i$ & $j$ & $k$ & $-1$ & $-i$ & $-j$ & $-k$\\ \hline\hline
$1$ & $1$ & $i$ & $j$ & $k$ & $-1$ & $-i$ & $-j$ & $-k$\\ \hline
$i$ & $i$ & $-1$ & $k$ & $-j$ & $-i$ & $1$ & $-k$ & $j$\\ \hline
$j$ & $j$ & $-k$ & $-1$ & $i$ & $-j$ & $k$ & $1$ & $-i$\\ \hline
$k$ & $k$ & $j$ & $-i$ & $-1$ & $-k$ & $-j$ & $i$ & $1$\\ \hline
$-1$ & $-1$ & $-i$ & $-j$ & $-k$ & $1$ & $i$ & $j$ & $k$\\ \hline
$-i$ & $-i$ & $1$ & $-k$ & $j$ & $i$ & $-1$ & $k$ & $-j$\\ \hline
$-j$ & $-j$ & $k$ & $1$ & $-i$ & $j$ & $-k$ & $-1$ & $i$\\ \hline
$-k$ & $-k$ & $-j$ & $i$ & $1$ & $k$ & $j$ & $-i$ & $-1$\\ \hline
\end{tabular}
\label{multgrupo}
\end{table}

\paragraph{The Klein group $V$.}
$V=\{e,a,b,ab\}$ is abelian, with each element self-inverse. We use it as the symmetry group of a planar frame (Table~\ref{Kleinpro}; \citealp{fraleigh1990linear,alexandroff2012introduction}).

\begin{table}[h]
    \centering
    \caption{Cayley table for the Klein four–group $\mathbb{V}_4=\{e,a,b,ab\}$, included for comparison with the quaternion group.}
    \begin{tabular}{|c||c|c|c|c|}
    \hline
    $\ast$ & $e$ & $a$ & $b$ & $ab$\\ \hline\hline
    $e$ & $e$ & $a$ & $b$ & $ab$\\ \hline
    $a$ & $a$ & $e$ & $ab$ & $b$\\ \hline
    $b$ & $b$ & $ab$ & $e$ & $a$\\ \hline
    $ab$ & $ab$ & $b$ & $a$ & $e$\\ \hline
    \end{tabular}
    \label{Kleinpro}
\end{table}

\begin{figure}[h]
\centering
\includegraphics[width=5cm,height=5cm]{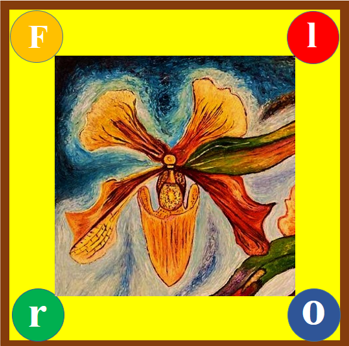}\hspace{1cm}
\includegraphics[width=5cm,height=5cm]{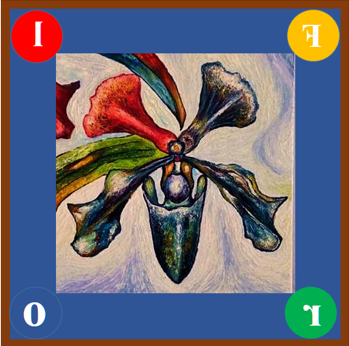}
\caption{Front (left) and back (right) views of the painting used in the didactic model.}
\label{cuadroinic}
\end{figure}

\begin{figure}[h]
\centering
\includegraphics[width=5cm,height=5cm]{Ident-I.png}\hspace{0.5cm}
\includegraphics[width=5cm,height=5cm]{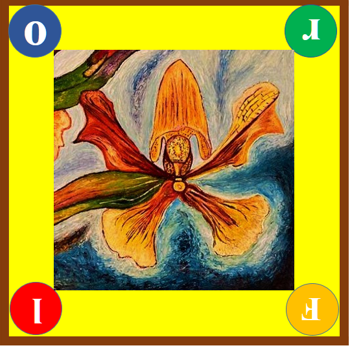}\\[6pt]
\includegraphics[width=5cm,height=5cm]{Ref-Fv.png}\hspace{0.5cm}
\includegraphics[width=5cm,height=5cm]{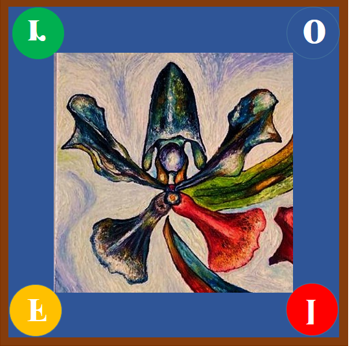}
\caption{Transformations of the painting identified with the Klein four-group $V$: (a) initial position $I$; (b) a $\pi$ rotation about the out-of-plane axis $R_2$; (c) vertical reflection $f_v$; (d) horizontal reflection $f_h$.}
\label{rotcuadroinic}
\end{figure}

\paragraph{The initial frame (IF) and $V$.}
Identify $e$ with the initial frame $I$, $a$ with a $\pi$ rotation about the out-of-plane axis ($R_2$), $b$ with the horizontal flip $f_h$, and $ab$ with the vertical flip $f_v$ (Figs.~\ref{cuadroinic}, \ref{rotcuadroinic}). These satisfy
\begin{equation*}
    R_2^2=f_h^2=f_v^2=I,\qquad R_2 f_h=f_v,\qquad f_h f_v=R_2.
\end{equation*}

\begin{figure}[h]
\centering
\includegraphics[width=5cm,height=7cm]{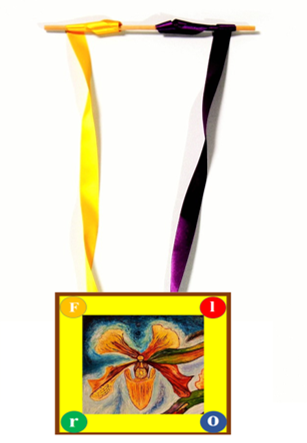}
\caption{Initial, untwisted state of the didactic setup: a painting suspended from two ribbons (yellow and violet) attached to a horizontal support. This configuration is identified with the identity element $1$ in the quaternion model.}
\label{cuadrocolg}
\end{figure}

\begin{figure}[h]
\centering
\includegraphics[width=6cm,height=5cm]{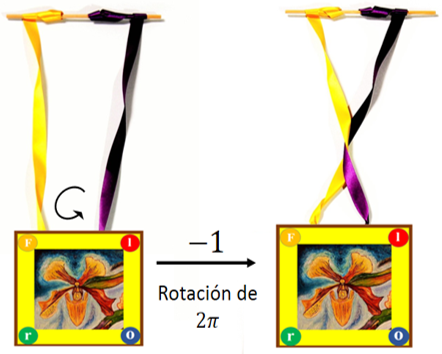}\hspace{1cm}
\includegraphics[width=6cm,height=5cm]{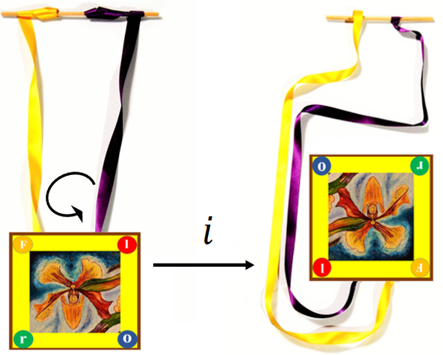}\\[10pt]
\includegraphics[width=7cm,height=5cm]{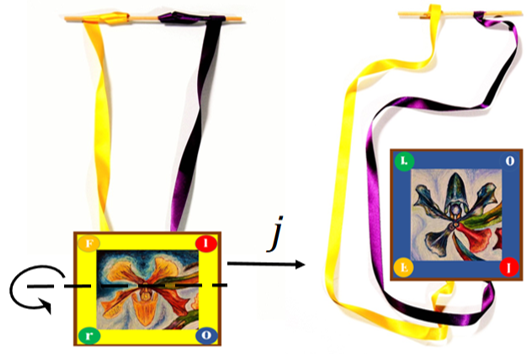}\hspace{1cm}
\includegraphics[width=6cm,height=5cm]{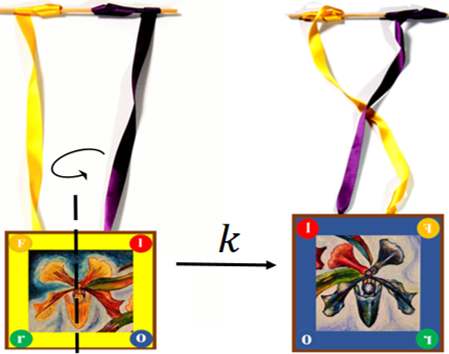}
\caption{Quaternion elements realised with the painting–with–strings model. (a) A $\pi$ rotation about the out-of-plane axis yields the configuration labelled $i$. (b) A $\pi$ rotation about the horizontal axis yields $j$. (c) A $\pi$ rotation about the vertical axis yields $k$. (d) A full $2\pi$ rotation returns the painting’s orientation but twists the ribbons, producing the state $-1$.}

\label{identicuadroinic}
\end{figure}

\paragraph{The frame with strings (FWS) and $\mathbb Q$.}
To model $\mathbb Q$ we need strings attached to the frame (the “puppet” of \citealp{diaz2017fenomeno,kauffman2001knots}). Hung by two strings (yellow and violet), the initial state is identified with $1\in\mathbb Q$ (Fig.~\ref{cuadrocolg}). A $2\pi$ rotation about the vertical axis returns the image but tangles the strings: we identify this with $-1$. Rotations by $\pi$ about three orthogonal axes are identified with $i$, $j$ and $k$ (Fig.~\ref{identicuadroinic}). Consequently,
\[
i^2=j^2=k^2=ijk=-1,
\]
and full untangling requires a total of $4\pi$ (Dirac belt trick) \citep{silverman1980curious,staley2010understanding,bolker1973spinor,penrose1984spinors,hansen1994magic,stojanoska2008touching}.

\begin{figure}[h]
\centering
\includegraphics[width=3.5cm,height=7cm]{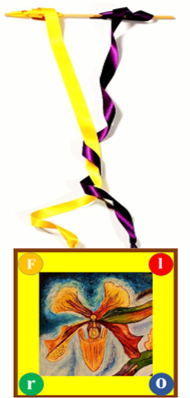}\hspace{0.5cm}
\includegraphics[width=4cm,height=7cm]{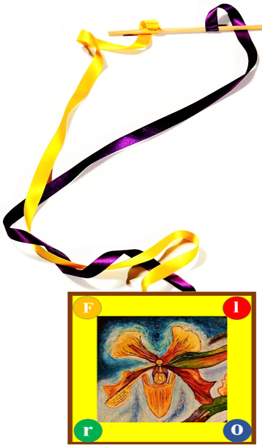}\hspace{0.5cm}
\includegraphics[width=4cm,height=7cm]{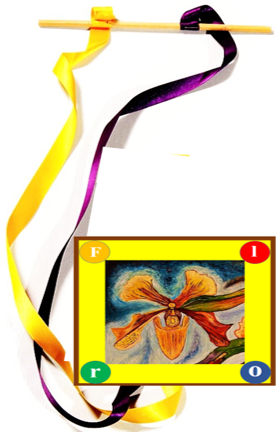}\\[10pt]
\includegraphics[width=4cm,height=7cm]{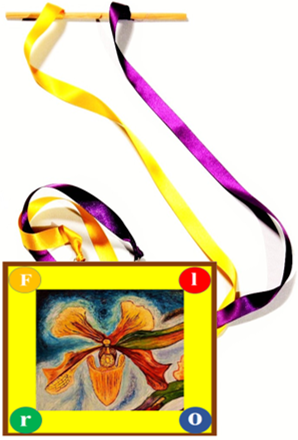}\hspace{0.5cm}
\includegraphics[width=3.5cm,height=7cm]{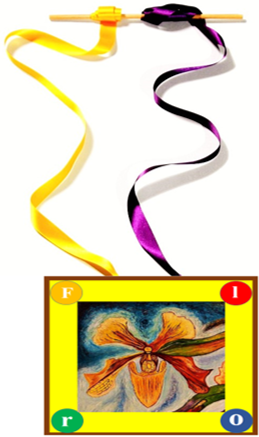}\hspace{1cm}
\includegraphics[width=3.5cm,height=7cm]{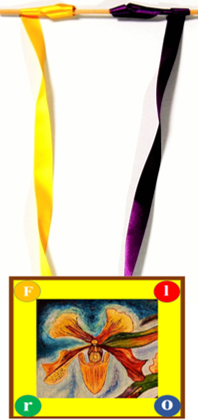}
\caption{Untangling sequence after a total $4\pi$ rotation. Starting from the twisted $2\pi$ state, the ribbons are gradually re-routed around the suspension bar while the painting remains fixed. The final configuration reproduces the initial, untwisted state, visually demonstrating the spinorial $4\pi$ periodicity.}

\label{desenredoCUIC}
\end{figure}

\paragraph{Untangling ($4\pi$).}
After a $2\pi$ turn the strings are tangled ($-1$). A second $2\pi$ (total $4\pi$) allows untangling while keeping the frame fixed and moving only the strings (Fig.~\ref{desenredoCUIC}); in group terms, $(-1)^2=1$.

\begin{figure}[h]
\centering
\includegraphics[width=6cm,height=6cm]{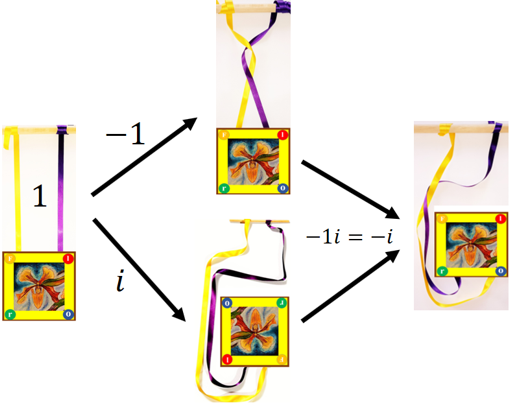}\hspace{1cm}
\includegraphics[width=6cm,height=8cm]{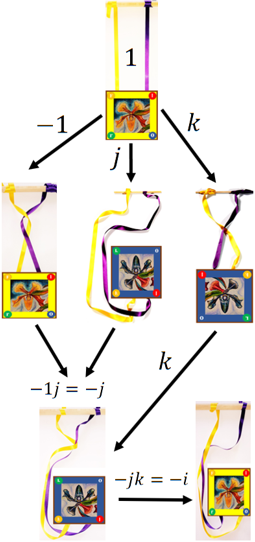}
\caption{Composition of quaternionic rotations in the painting–with–strings model. (a) Subgroup generated by $\{1,i,-1,-i\}$ showing the $i^2=-1$ relation through successive $\pi$ rotations about the same axis (left panel). (b) Extended composition involving $j$ and $k$, illustrating noncommutativity and the identities $jk=-i$ and $k^2=-1$. The visual mapping of tangled and untangled states reproduces the group structure of $\mathbb Q_8$ (rigth panel).}

\label{ejemplo-i}
\end{figure}

\paragraph{Examples (products in $\mathbb Q$ as rotation sequences).}
The configuration $-i$ equals a $2\pi$ twist (state $-1$) followed by the rotation defining $i$ (a $\pi$ turn about the out-of-plane axis). Likewise, $(-j)k=-i$ corresponds to a $2\pi$ twist plus the horizontal flip ($j$) and then the vertical flip ($k$), see Fig.~\ref{ejemplo-i}.

\subsection*{Quaternions and Pauli Matrices}

Let
\[
L=
\begin{bmatrix}
a+d & b-ic\\
b+ic & a-d
\end{bmatrix},\qquad a,b,c,d\in\mathbb R,
\]
a Hermitian matrix with $\det L=a^2-b^2-c^2-d^2$. Define $f:\mathbb R^4\!\to\! ML$ by $f(a,b,c,d)=L$ (linear). On the canonical basis:
\[
f(e_0)=\sigma_0,\quad f(e_1)=\sigma_x,\quad f(e_2)=\sigma_y,\quad f(e_3)=\sigma_z.
\]
Multiplying by $i$, the space $P=\{\sigma_0,i\sigma_x,i\sigma_y,i\sigma_z\}$ is isomorphic to $\mathbb H$ via
\[
h(a+bi+cj+dk)=a\sigma_0 - i(b\sigma_x+c\sigma_y+d\sigma_z),
\]
with linearity and product-preserving properties (see the main text for details).

Now, for $q=a+bi+cj+dk$ with $|q|=1$, define
\[
g(q)=
\begin{bmatrix}
d+ic & b+ia\\
-b+ia & d-ic
\end{bmatrix}.
\]
Then $\det g(q)=1$ and $g(q)$ is unitary: $g(q)(g(q)^{\ast})^{T}=I$. Hence,
\[
g:S^3\longrightarrow SU(2)
\]
is a group isomorphism; this articulates the familiar correspondence $S^3\simeq SU(2)$ and explains the $S^3\to SO(3)$ double cover underlying the global removal of singularities.

\end{document}